\documentclass[12pt, a4paper]{article}
\usepackage{epsf}
\usepackage{cite}
\usepackage{amsmath,amssymb}
\input{colordvi.tex}
\usepackage[usenames,dvipsnames]{color}
\usepackage[dvips]{graphicx}
\usepackage{comment}
\begin{document}

\newcommand{\lsim}{\stackrel{<}{_\sim}}
\newcommand{\gsim}{\stackrel{>}{_\sim}}
\newcommand{\rem}[1]{{ {\color{red} [[$\spadesuit$ \bf #1 $\spadesuit$]]} }}

\renewcommand{\theequation}{\thesection.\arabic{equation}}
\renewcommand{\thefootnote}{\fnsymbol{footnote}}
\setcounter{footnote}{0}


\def\thefootnote{\fnsymbol{footnote}}
\def\a{\alpha}
\def\b{\beta}
\def\c{\varepsilon}
\def\d{\delta}
\def\e{\epsilon}
\def\f{\phi}
\def\g{\gamma}
\def\h{\theta}
\def\k{\kappa}
\def\l{\lambda}
\def\m{\mu}
\def\n{\nu}
\def\p{\psi}
\def\q{\partial}
\def\r{\rho}
\def\s{\sigma}
\def\t{\tau}
\def\u{\upsilon}
\def\v{\varphi}
\def\w{\omega}
\def\x{\xi}
\def\y{\eta}
\def\z{\zeta}
\def\D{\Delta}
\def\G{\Gamma}
\def\H{\Theta}
\def\L{\Lambda}
\def\F{\Phi}
\def\P{\Psi}
\def\S{\Sigma}
\def\me{\mathrm e}

\def\o{\over}
\def\beq{\begin{eqnarray}}
\def\eeq{\end{eqnarray}}
\newcommand{\vev}[1]{ \left\langle {#1} \right\rangle }
\newcommand{\bra}[1]{ \langle {#1} | }
\newcommand{\ket}[1]{ | {#1} \rangle }
\newcommand{\bs}[1]{ {\boldsymbol {#1}} }
\newcommand{\mc}[1]{ {\mathcal {#1}} }
\newcommand{\mb}[1]{ {\mathbb {#1}} }
\newcommand{\EV}{ {\rm eV} }
\newcommand{\KEV}{ {\rm keV} }
\newcommand{\MEV}{ {\rm MeV} }
\newcommand{\GEV}{ {\rm GeV} }
\newcommand{\TEV}{ {\rm TeV} }
\def\diag{\mathop{\rm diag}\nolimits}
\def\Spin{\mathop{\rm Spin}}
\def\SO{\mathop{\rm SO}}
\def\O{\mathop{\rm O}}
\def\SU{\mathop{\rm SU}}
\def\U{\mathop{\rm U}}
\def\Sp{\mathop{\rm Sp}}
\def\SL{\mathop{\rm SL}}
\def\tr{\mathop{\rm tr}}
\def\sp{\;\;}

\def\IJMP{Int.~J.~Mod.~Phys. }
\def\MPL{Mod.~Phys.~Lett. }
\def\NP{Nucl.~Phys. }
\def\PL{Phys.~Lett. }
\def\PR{Phys.~Rev. }
\def\PRL{Phys.~Rev.~Lett. }
\def\PTP{Prog.~Theor.~Phys. }
\def\ZP{Z.~Phys. }

\begin{titlepage}

\begin{center}

\hfill UT-15-15\\
\hfill IPMU-15-0056\\
\hfill April, 2015\\

\vskip .75in

{\Large \bf 
Particle Production after Inflation with\\[.5em]
Non-minimal Derivative Coupling to Gravity
}

\vskip .75in

{\large Yohei Ema$^a$, Ryusuke Jinno$^a$, Kyohei Mukaida$^b$ and Kazunori Nakayama$^{a,b}$}

\vskip 0.25in

\begin{tabular}{ll}
$^{a}$&\!\! {\em Department of Physics, Faculty of Science, }\\
& {\em The University of Tokyo,  Bunkyo-ku, Tokyo 133-0033, Japan}\\[.3em]
$^{b}$ &\!\! {\em Kavli IPMU (WPI), UTIAS,}\\
&{\em The University of Tokyo,  Kashiwa, Chiba 277-8583, Japan}
\end{tabular}

\end{center}
\vskip .5in

\begin{abstract}
We study cosmological evolution after inflation in models with non-minimal derivative coupling to gravity.
The background dynamics is solved and particle production associated with rapidly oscillating Hubble parameter is studied in detail.
In addition, production of gravitons through the non-minimal derivative coupling with the inflaton is studied.
We also find that the sound speed squared of the 
scalar perturbation oscillates between positive and negative values 
when the non-minimal derivative coupling dominates over the minimal kinetic term. This may lead to 
an instability of this model. We point out that the particle production rates are the same as those 
in the Einstein gravity with the minimal kinetic term, if we require the sound speed squared is positive definite.
\end{abstract}

\end{titlepage}

\tableofcontents

\renewcommand{\thepage}{\arabic{page}}
\setcounter{page}{1}
\renewcommand{\thefootnote}{\#\arabic{footnote}}
\setcounter{footnote}{0}

\newpage

\section{Introduction}
\setcounter{equation}{0}

Inflation theories, now accepted as a successful way of explaining the isotropy and homogeneity 
as well as the density fluctuation of the universe, assume the existence of a scalar field, or inflaton, 
to trigger inflation.
The coupling between the inflaton and gravity might in general be non-canonical, and 
among such theories there exists a class which introduces no extra degrees of freedom. 
This class of models has been investigated especially after the discovery of Higgs boson at 
LHC\cite{Aad:2012tfa,Chatrchyan:2012ufa} and after the results from Planck satellite\cite{Ade:2015lrj}, 
in the context of having successful inflation with the standard model (SM) Higgs field. 
For example, it was proposed to couple the Higgs field non-minimally to the Ricci scalar $R$\cite{Futamase:1987ua,
CervantesCota:1995tz,Bezrukov:2007ep}, 
or to introduce a non-minimal derivative coupling between the Higgs field and the Einstein tensor 
$G_{\mu \nu}$\cite{Germani:2010gm}.
Moreover, it was pointed out that a modification of the Higgs kinetic term from the canonical one 
also leads to a successful inflation\cite{Takahashi:2010ky}. 
For general setups for inflation using non-minimal couplings 
without introducing additional degrees of freedom, see the studies of the generalized G-inflation\cite{Kobayashi:2011nu}.

While inflation itself is a matter of great interest, the reheating phase is equally interesting 
not only because it necessarily follows the inflationary phase in order to produce SM particles,
but also because it is expected to have rich phenomenology.
To study the reheating phase, we must know the background evolution in the inflaton oscillation dominated universe and also how the inflaton couples to SM fields. 
In models with extended gravity, these aspects are often non-trivial compared with the standard inflation models with Einstein gravity.
Interestingly, it is sometimes possible to reheat the universe 
without introducing explicit interaction terms among the inflaton and SM fields,
since the inflaton necessarily couples to SM sector gravitationally
at the tree level~\cite{Arbuzova:2011fu,Watanabe:2006ku,Ema:2015dka}.

In this paper, we study the reheating in the model with non-minimal derivative coupling: $\mathcal L \sim G^{\mu \nu}
\partial_\mu \phi \partial_\nu \phi$.
We do not necessarily assume that $\phi$ is the SM Higgs field. 
The detailed analysis of inflaton oscillation dominated universe in this class of model was performed in Ref.~\cite{Jinno:2013fka}.
There it was found that the Hubble parameter rapidly oscillates if the non-minimal kinetic term of the inflaton takes a dominant role.
Thus it is expected that particle production of non-Weyl invariant light fields may be efficient.
More interestingly, the inflaton has a large non-minimal coupling to graviton,
which opens up a new possibility to produce significant amount of gravitons from the inflaton coherent oscillation.
Actually we will find that gravitons may be resonantly produced after inflation.
However, we will also point out that the sound speed squared of the 
scalar perturbation rapidly oscillates between positive and negative values 
in such a case and it may lead to an instability for the shortest wavelength mode of the scalar perturbation.
  
The organization of this paper is as follows. 
First we write down background equations of motion, 
and summarize different phases the universe undergoes in this model. 
In Sec.~\ref{sec:perturbation}, we study the perturbation around the background and find that the 
sound speed squared of the scalar perturbation can be negative.
In Sec.~\ref{sec:particle}, we analyze particle production in this model, 
focusing on minimally-coupled scalars and gravitons. 
Sec.~\ref{sec:conclusion} is devoted to conclusion.

\section{Background dynamics}
\label{sec:background}
\setcounter{equation}{0}

\subsection{Background equations}

 First of all, we derive the background equation~\cite{Germani:2010gm}.
We consider the following action
\begin{align}
S
&= \int d^4x \sqrt{-g}\left[\frac{M_P^2}{2}R - \frac{1}{2} \left(g^{\mu \nu} - \frac{G^{\mu \nu}}{M^2} \right)\partial_\mu \phi \partial_\nu \phi 
- V(\phi) \right],
\label{eq:action}
\end{align}
where $\phi$ is a real scalar field, or the inflaton, $R$ is the Ricci scalar and $G_{\mu \nu} = R_{\mu \nu} - g_{\mu \nu}R/2$ is the Einstein tensor. 
Also, $M_P = (8\pi G)^{-1/2}$ is the reduced Planck mass and $M$ is some mass parameter.
We assume the background part of the metric to be the Friedmann-Lema\^itre-Robertson-Walker (FLRW) one with negligible curvature
\begin{align}
ds^2
&= -dt^2 + a(t)^2dx^idx^i,
\end{align}
where $a$ is the scale factor.
After integration by parts, the background action becomes
\begin{align}
S
&= \int d^4x \; a^3 \left[ -3M_P^2H^2 + \left( 1 + \frac{3H^2}{M^2} \right)\frac{\dot{\phi}^2}{2} - V(\phi) \right],
\label{eq:bgaction}
\end{align}
where $H = \dot{a}/a$ is the Hubble parameter.

The Friedmann equation is obtained by introducing the lapse function $N$ by $dt \rightarrow Ndt$ and taking 
variation with respect to it:
\begin{align}
&H^2
= \frac{\rho_\phi}{3M_P^2}, 
\label{eq:Friedmann}\\
&\rho_\phi
\equiv \left( 1+ \frac{9H^2}{M^2} \right)\frac{\dot{\phi}^2}{2} + V,
\label{eq:rhophi}
\end{align}
where $\rho_{\phi}$ is the energy density of $\phi$ and we have taken $N = 1$ after the variation.
Note that the Friedmann equation~\eqref{eq:Friedmann} implies the following relation:
\begin{align}
\epsilon
&\equiv \frac{\dot{\phi}^{2}}{M_{P}^{2}M^{2}} < \frac{2}{3}.
\label{eq:expansion}
\end{align}
The variation with respect to the scale factor gives
\begin{align}
&3H^2 + 2\dot{H}
= -\frac{p_\phi}{M_P^2},
\label{eq:pressure}\\
&p_\phi
\equiv \left( 1 - \frac{3H^2}{M^2} \right) \frac{\dot{\phi}^2}{2} - V - \frac{1}{M^2}\frac{d}{dt}(H\dot{\phi}^2),
\label{eq:pphi}
\end{align}
where $p_{\phi}$ is the pressure of $\phi$. As we will see later, violent oscillating features of 
$\phi$'s energy density
and of the Hubble parameter for $H \gg M$ originate from the last term in Eq.~\eqref{eq:pphi}.
The equation of motion for $\phi$ is
\begin{align}
\left( 1 + \frac{3H^2}{M^2} \right)\ddot{\phi} + 3H\left( 1 + \frac{3H^2}{M^2} + \frac{2\dot{H}}{M^2} \right)\dot{\phi} 
+ \frac{\partial V}{\partial \phi} 
&= 0.
\label{eq:phiEOM}
\end{align}
From this equation of motion, we can see that the potential is effectively suppressed due to the non-minimal
derivative coupling for $H/M \gg 1$.
Note that Eq.~\eqref{eq:phiEOM} is equivalent to the energy conservation equation
\begin{align}
\dot{\rho_{\phi}} + 3H\left(\rho_{\phi} + p_{\phi}\right)
&= 0,
\label{eq:rhodot}
\end{align}
and that one of the three equations~\eqref{eq:Friedmann}, \eqref{eq:pressure} and \eqref{eq:phiEOM} is redundant.
Also, $\dot{H}$ is calculated by eliminating $\ddot{\phi}$ from the time derivative of Eq.~(\ref{eq:Friedmann}) and 
Eq.~(\ref{eq:phiEOM}) as
\begin{align}
\frac{\dot{H}}{M^2}
&= -\frac{(1+\frac{3H^2}{M^2})(1+\frac{9H^2}{M^2})\frac{\epsilon}{2} + \frac{HV'}{M^2 \dot{\phi}}\epsilon}
{(1+ \frac{3H^2}{M^2}) - (1-\frac{9H^2}{M^2})\frac{\epsilon}{2}},
\label{eq:Hdot}
\end{align}
where the prime denotes the derivative with respect to $\phi$.

In the following, we consider the power-law potential
\begin{align}
V
&= \frac{\lambda}{n}\phi^n.
\end{align}
Here we define the effective mass $m_{\rm eff}$ as
\begin{align}
m_{\rm eff}
&\equiv
\left\{
\begin{matrix}
\frac{M}{H} \left. \sqrt{ \frac{V'}{\phi} } \right|_{\phi \rightarrow \Phi} 
&= \frac{M}{H}\sqrt{\lambda}\Phi^{\frac{n}{2}-1}
\;\;\;(\frac{H}{M} \gg 1) \\
\left. \sqrt{ \frac{V'}{\phi} } \right|_{\phi \rightarrow \Phi} 
&= \sqrt{\lambda}\Phi^{\frac{n}{2}-1}
\;\;\;\;\;\;\;\;\;(\frac{H}{M} \ll 1) \\
\end{matrix}
\right.
\label{eq:meff}
\end{align}
where $\Phi$ is the amplitude of the $\phi$ oscillation. Note that the effective mass is an increasing
function with time due to the suppression factor $M/H$ for $H/M \gg 1$. 
Also note that $m_\text{eff}  \Phi$ remains constant for $H/M \gg 1$
in the inflaton oscillation dominated era.
We sometimes express
$\sqrt{\lambda}$ as $m_{\phi}$ for $n=2$.

\subsection{Evolution of the universe}

The evolution of the universe for models with the non-minimal derivative 
coupling was investigated in detail in Ref.~\cite{Jinno:2013fka}. Here, we briefly summarize the discussion given 
there.
We assume that the universe is dominated by the inflaton $\phi$ and neglect the damping of $\phi$ due to particle production in this subsection.

First, consider the case where $m_{\rm{eff}} \ll H$ (Phase~0). In this case, inflation takes place due to the large
Hubble friction term. We summarize the inflationary predictions of the models with non-minimal derivative coupling
in App.~\ref{app:perturbation}.

Next, consider the case where $M \ll H \ll m_{\rm{eff}}$ (Phase~1). In this case, $\phi$ starts to oscillate
while the kinetic term is dominated by the non-minimal one. 
In analyzing this phase, we note that the energy
density of $\phi$ oscillates violently with time. To see this, one may refer to Eq.~\eqref{eq:rhodot}, which is given 
after substituting the explicit forms of $\rho_{\phi}$ and $p_{\phi}$ as
\begin{align}
\frac{d}{dt}\left[\left(1+\frac{9H^2}{M^2}\right)\frac{\dot{\phi}^2}{2} + V \right] 
+ 3H\left(1+\frac{3H^2}{M^2}\right)\dot{\phi}^2 - \frac{3H}{M^2}\frac{d}{dt}\left(H\dot{\phi}^2\right)
&= 0.
\label{eq:rho_explicit}
\end{align}
The second term is $\mathcal{O}(H\rho_{\phi})$, and no oscillating feature is 
caused by this term. On the other hand, the third term is $\mathcal{O}(m_{\rm{eff}}\rho_{\phi})$, 
and hence $\rho_{\phi}$ oscillates with the frequency of $\sim m_{\rm{eff}}$ due to this term.
Note that it means that the Hubble parameter also violently oscillates with time when $\phi$ dominates the universe.
Because of this oscillating feature, $\rho_{\phi}$ does not suit for the evaluation of the evolution of the universe 
in this regime.
Instead, it is shown in Ref.~\cite{Jinno:2013fka} that the following quantity $J$ is an adiabatic invariant:\footnote{
Throughout this paper we call a quantity $I$ an adiabatic invariant if $I$ satisfies $\dot{I} \sim \mathcal{O}(HI)$.}
\begin{align}
J &\equiv \frac{1}{H} \left[ \left( 1 + \frac{6H^2}{M^2} \right)\frac{\dot{\phi}^2}{2} + V \right]
= 3M_P^2H\left(1 - \frac{1}{2}\frac{\dot{\phi}^2}{M_P^2M^2}\right).
\end{align}
We prove the adiabaticity of $J$ in App.~\ref{app:adiabaticity}.
We can see from this equation that the oscillation of the Hubble parameter and that of $\dot{\phi}^2$ correlate with each other.    
In Ref.~\cite{Jinno:2013fka}, $J$ is extensively used to calculate the expansion law of the universe in this regime. 
The expansion law in this regime was derived as
\begin{align}
	\left< H\right> = \frac{2n+2}{3n}\frac{1}{t},
\end{align}
where the parenthesis denotes the time average.
We further recall the important feature
in Phase~1 here.
Notice that the Friedmann equation~\eqref{eq:Friedmann} implies that 
\begin{align}
\frac{m_{\rm{eff}}^{2}\Phi^{2}}{M_{P}^{2}M^{2}} \sim \mathcal{O}(1),
\label{eq:qphase1}
\end{align}
is always kept in Phase~1 when $\phi$ dominates the universe. This is because the amplitude $\Phi$ decreases while
the effective mass $m_{\rm{eff}}$ increases with time, and hence $m_{\rm{eff}}\Phi$ is almost constant during Phase~1. 
This feature will become important
when we discuss particle production in Sec.~\ref{sec:particle}.

Finally, consider the case where $H \ll M$ (Phase~2). In this case, the kinetic term of $\phi$ is dominated by the 
minimal one, and the evolution of the universe approaches to the ordinary case of the minimal oscillating scalar with Einstein gravity.
The expansion law in this case is given by\footnote{
In fact, for $M(M/m_{\rm{eff}})^{1/3} \ll H \ll M$, 
the term proportional to $\dot\phi$ is still dominated by the term $\sim \dot{H}/M^{2}$ in Eq.~\eqref{eq:phiEOM}.
However, this term does not act as friction because it is not positive definite, 
and the expansion law soon becomes the ordinary one described by Eq.~(\ref{H_Ein}) after a few Hubble time after entering the phase 2.
}
\begin{align}
	H = \frac{n+2}{3n}\frac{1}{t}.  \label{H_Ein}
\end{align}
Whether Phase 1 exists or not depends on the model parameters. 
In order to discuss each situation separately, we consider the following two cases:
\begin{itemize}
\item Case~A : Phase 1 does not exist. This corresponds to the case where $H < M$ 
and $m_{\rm eff} < M$ are always satisfied after inflation. It requires 
\begin{equation}
	\frac{\lambda M_P^{n-2}}{M^2} \lesssim 1,   \label{nophase1}
\end{equation}
(see Eq.~\eqref{nonmin_cond}). 
Then, the cosmic evolution is the same as that of the minimal scalar field with Einstein gravity.
Note that we always have $\dot H\sim H^2$ in this case.

\item Case~B : Phase 1 exists. In this case, the inequality (\ref{nophase1}) is inverted.
Then, as we discussed just above, the universe undergoes the following phases if we neglect
particle production:
\begin{itemize}
\item{Phase~0} : Inflation takes place for
\begin{align}
m_{\rm eff} \ll H
\;\leftrightarrow\;
M_P \left( \frac{M^2}{\lambda M_P^{n-2}} \right)^{\frac{1}{n+2}}
\ll \Phi.
\label{eq:HphiPhase0}
\end{align}
Due to the non-minimal kinetic term, the slow-roll conditions are easier to be satisfied than usual.
Note that this condition is also the same as $M_P \ll H\Phi/M$.

\item{Phase~1}
: $\phi$ starts to oscillate while the kinetic term is dominated by the non-minimal one for
\begin{align}
M \ll H \ll m_{\rm eff}
\label{eq:HphiPhase1}
\end{align}
In this regime, the energy density $\rho_{\phi}$ violently oscillates with time. When $\phi$ dominates the 
universe, this means that the Hubble parameter also violently oscillates with time.
Notice that $m_{\rm{eff}}$ is an increasing function with time in this regime.

\item{Phase~2}
: The kinetic term of $\phi$ in Eq.~(\ref{eq:Friedmann}) is dominated by the standard one for
\begin{align}
 H \ll M
\label{eq:HphiPhase2}
\end{align}
In this regime, the evolution of the universe is the same as the ordinary case of the minimal oscillating scalar with Einstein gravity.

\end{itemize}

Note that the Hubble parameter can have a large oscillating part in this case. In fact, the behavior of $\dot{H}$ can be classified into the 
following three regimes:
\begin{equation}
	\dot H \sim \left \{ \begin{array}{ll}
	  m_{\rm eff} H& ~~{\rm for}~~ M<H ~~(\rm{ Phase~1}), \\
	  m_{\rm eff}H^3 / M^2& ~~{\rm for}~~M^2/m_{\rm eff} < H < M ~~(\rm{ Phase~2}),\\
	  H^2&~~{\rm for}~~H < M^2/m_{\rm eff} ~~(\rm{Phase~2}).
	\end{array}
	\right.
	\label{eq:dotHubble}
\end{equation}

\end{itemize}
The reason why we consider these two cases will become clear in the next section.
In the rest of the paper, we focus on the oscillation regime (Phases~1--2 for Case~B) unless otherwise stated.

\section{Perturbation}
\label{sec:perturbation}
\setcounter{equation}{0}

\subsection{Scalar perturbation}  \label{sec:scalarpert}

In this subsection we derive the quadratic action for the scalar perturbation when the inflaton $\phi$ dominates the universe. 
The calculation basically follows those done in Refs.~\cite{Germani:2010gm,Kobayashi:2011nu,Germani:2011ua},
although they focused on the inflationary regime.
Below we will see that the perturbed quantities exhibit qualitatively different behavior in the inflaton oscillating regime.

We must first solve the mixing between the scalar part of the metric and the inflaton. We use the Arnowitt-Deser-Misner (ADM)
formalism, in which the metric is taken as
\begin{align}
ds^{2} = -N^{2}dt^{2} + \gamma_{ij}\left(dx^{i} + \beta^{i}dt\right)\left(dx^{j}+\beta^{j}dt\right),
\end{align}
where $N$ is the lapse function, $\beta^{i}$ is the shift vector and $\gamma_{ij}$ is the 3-dimensional 
spatial metric. 
We take the unitary gauge $\phi(t, \vec{x}) = \phi(t)$ in the calculation. Then, the action is given by
\begin{align}
S = \int dx^{4}\sqrt{-g}\left[\frac{M_{P}^{2}}{2}R - \left(g^{00}-\frac{G^{00}}{M^{2}}\right)\frac{\dot{\phi}^{2}}{2} - V\right].
\end{align}
In the ADM formalism, the Ricci scalar and the (00)-component of the Einstein tensor are decomposed as
\begin{align}
R &= R^{(3)} + \frac{1}{N^{2}}\left(E^{ij}E_{ij} - E^{2}\right) - 2 \nabla_{\mu}\left(Kn^{\mu}\right) - \frac{2}{N}\Delta^{(3)}N, \\
G^{00} &= \frac{1}{2N^{2}}\left(R^{(3)} + \frac{1}{N^{2}}\left(E^{2} - E^{ij}E_{ij}\right)\right),
\end{align}
where $E_{ij}$ is related to the extrinsic curvature $K_{ij}$ as
\begin{align}
E_{ij} &= NK_{ij} = \frac{1}{2}\left(\nabla_{i}^{(3)}\beta_{j} + \nabla_{j}^{(3)}\beta_{i} - \dot{\gamma}_{ij}\right),
\end{align}
the superscript $``(3)"$ denotes covariant quantities with respect to the spatial metric $\gamma_{ij}$
and $\gamma_{ij}$ is used to take the summation with respect to the space indices $i, j, ...$. 
Moreover, $n^{\mu}$ is the unit normal vector of the time-like hyper-surface, whose components are given by
\begin{align}
	n^{\mu} = \frac{1}{N}\begin{pmatrix}
	1, &-\beta^{i} \\
\end{pmatrix}.
\end{align}
Thus, we rewrite the action as
\begin{align}
S 
= \int dx^{4} \frac{M_{P}^{2}\sqrt{\gamma}}{2}
&
\left[R^{(3)}\left(N + \frac{\dot{\phi}^{2}}{2NM_{P}^{2}M^{2}}\right) 
\right. \nonumber \\
&+ \left. \left(E^{ij}E_{ij} - E^{2}\right)\left(\frac{1}{N} 
- \frac{\dot{\phi}^{2}}{2N^{3}M_{P}^{2}M^{2}}\right) 
+ \frac{\dot{\phi}^{2}}{NM_{P}^{2}} - \frac{2NV}{M_{P}^{2}}\right].
\end{align}
Note that the action does not include time derivatives of the lapse function and the shift vector. 
Therefore the derivatives of the action with respect to them yield constraint equations, 
which can be used to relate $N$ and $\beta^{i}$ to the other scalar perturbations. 
The constraint equation for $N$ is
\begin{align}
R^{(3)}\left(1-\frac{\dot{\phi}^{2}}{2N^{2}M_{P}^{2}M^{2}}\right) 
- \left(E^{ij}E_{ij} - E^{2}\right)\left(\frac{1}{N^{2}}-\frac{3\dot{\phi}^{2}}{2N^{4}M_{P}^{2}M^{2}}\right)
- \frac{\dot{\phi}^{2}}{N^{2}M_{P}^{2}} - \frac{2V}{M_{P}^{2}} = 0,
\label{eq:lapse}
\end{align}
and that for $\beta^{i}$ is 
\begin{align}
\nabla_{i}^{(3)}\left[\left(\frac{1}{N}-\frac{\dot{\phi}^{2}}{2N^{3}M_{P}^{2}M^{2}}\right)\left(E^{i}_{\;j} - \delta^{i}_{\;j}E\right)\right] = 0.
\label{eq:shift}
\end{align}
Now, we expand the metric as\footnote{
Since the vector and tensor perturbations do not couple to the scalar perturbations in the lowest order, 
we neglect them here.}
\begin{align}
	N &= 1+\alpha, \nonumber \\
	\beta_{i} &= \partial_{i}\psi, \nonumber \\
	\gamma_{ij} &= a(t)^{2}e^{2\zeta}\delta_{ij},
	\label{eq:gauge}
\end{align}
where we have fixed the remaining one scalar degree of freedom 
by the gauge transformation.
To zeroth order in perturbations, Eq.~\eqref{eq:lapse} is nothing but the Friedmann equation Eq.~\eqref{eq:Friedmann}.
To first order, we obtain the following relations among $\alpha, \psi$ and $\zeta$ from Eqs.~\eqref{eq:lapse} and \eqref{eq:shift}:\footnote{
Although we write down the solution for $\psi$ here, $\psi$ does not contribute to the quadratic action 
because it appears linearly in the quadratic action after integration by parts.
}
\begin{align}
\alpha 
&= \frac{F}{H}\dot{\zeta}, \; \; \; \; 
\psi 
= -\frac{F}{H} \zeta + \chi, \; \; \; \; 
\partial_{i}^{2}\chi 
= a^2\frac{M^2}{H^2} \frac{F^2G}{1-\frac{1}{2}\epsilon}\dot{\zeta},
\end{align}
where
\begin{align}
F 
&= \frac{1-\frac{1}{2}\epsilon}{1-\frac{3}{2}\epsilon}, \; \; \; \; 
G
= \frac{\epsilon}{2}\left(1+\frac{3H^{2}}{M^{2}}\frac{1+\frac{3}{2}\epsilon}{1-\frac{1}{2}\epsilon}\right).
\end{align}
See Eq.~\eqref{eq:expansion} for the definition of $\epsilon$.
By substituting these relations to the original action and performing some integration by parts, we obtain 
the following quadratic action for $\zeta$:\footnote{
At the end-points of the oscillation, this action may be ill-defined since $\dot{\phi}=0$ at those points. 
However, the expression of the sound speed squared is the same in other gauges as
we have shown in App.~\ref{app:delphi_gauge},} and hence
the following discussion is not affected by this subtlety.
\begin{align}
S = M_{P}^{2}M^2\int dx^{4}\; a^{3}\frac{F^{2}G}{H^{2}}\left[\dot{\zeta}^{2} - \frac{c_{s}^{2}}{a^{2}}\left(\partial_{i}\zeta\right)^{2}\right],
\label{eq:quad-action}
\end{align}
where $c_{s}$ is the sound speed, which is given by
\begin{align}
c_{s}^{2}
&=\frac{1}{K}
\left[\left(1+\frac{3}{2}\epsilon\right) 
+ \frac{3H^2}{M^2}\left[ \left(1+\frac{3}{2}\epsilon\right) 
+ \frac{2}{3F}\epsilon \right]
+ \frac{6\dot{H}}{M^2}\left(1-\frac{1}{2}\epsilon\right)\right], 
\label{eq:sound} \\
K
&\equiv \left(1-\frac{1}{2}\epsilon\right) \left(1+\frac{3H^2}{M^2}\frac{1+\frac{3}{2}\epsilon}{1-\frac{1}{2}\epsilon}\right),
\end{align}
where we have used Eq.~\eqref{eq:pressure} to express $\ddot{\phi}$ in terms of $\dot{H}$.
Note that due to Eq.~\eqref{eq:expansion}, $F^{2}G/H^{2}$ in Eq.~\eqref{eq:quad-action}
is always positive and no ghost mode exists
in the scalar perturbation. 
For the sound speed squared $c_{s}^{2}$, however, the situation is different.
The term proportional to $\dot{H}/M^{2}$ in the numerator in Eq.~\eqref{eq:sound} can be positive or negative due to the violent oscillating feature of the 
Hubble parameter, and hence $c_{s}^{2}$ can be negative if this term dominates over the other terms.
We consider Cases~A and B separately.
\begin{itemize}

\item Case~A : In this case, one always has $H < M$ and $\dot{H} \sim H^{2}$.
Then we estimate the sound speed squared as
\begin{align}
c_{s}^{2} = 1 + \mathcal{O}\left(\frac{H^{2}}{M^{2}}\right),
\end{align}
where we have used $\dot{\phi}^{2}/M_{P}^{2}M^{2} \sim H^{2}/M^{2}$ for $H < M$.
Therefore, the sound speed squared $c_{s}^{2}$ is positive definite and no violently growing
mode exists in the scalar perturbation.

\item Case~B : From Eq.~\eqref{eq:dotHubble}, the sound speed squared in Phase~1 is given as
\begin{align}
c_{s}^{2} \sim \frac{\dot{H}}{H^{2}}~~{\rm{for}}~~M<H.
\end{align}
Even in Phase~2, the term proportional to $\dot{H}/M^{2}$ dominates over the other terms for
$M(M/m_{\rm{eff}})^{1/3} < H < M$, and hence 
the sound speed squared is given as
\begin{align}
c_{s}^{2} \sim \frac{\dot{H}}{M^{2}}~~{\rm{for}}~~M(M/m_{\rm{eff}})^{1/3} < H < M.
\end{align}
Note that they are oscillating functions between positive and negative values in these regimes. 
This means that the sound speed squared becomes negative for some period during the oscillation.\footnote{
It is known that in some classes of the generalized Galileon theory
the sound speed squared can be negative in the oscillation regime\cite{Ohashi:2012wf}.
}
Hence, the scalar perturbation grows violently
and the theory has an instability.
In particular, the shortest wavelength modes are more likely to be enhanced, and hence we may need
some UV completion in order to discuss what really happens in this case. Such a discussion is beyond 
the scope of this paper. Thus we do not argue whether this is really problematic or not in this paper.

\end{itemize}
In summary, an instability due to the negative sound speed squared exists in the scalar perturbation
when the non-minimal derivative coupling dominates over the minimal kinetic term. If we require 
such an instability not to exist, the situation reduces to Case~A where there is no Phase~1.
This is why we consider both Cases~A and B in this paper.

\subsection{Tensor perturbation}
Next, we consider the action for the graviton.
We take the metric as
\begin{align}
ds^2 = -dt^2 + a^2(t)\left(e^{h}\right)_{ij}dx^{i}dx^{j},
\end{align}
where $h_{ij}$ is the transverse and traceless ($\partial_i h_{ij} = h_{ii} = 0$) part of the metric
and $e^{h}$ is the matrix exponential of $h$.
The relevant terms in the action are
\begin{align}
S_{\rm{grav}}
&= \int d^4x \sqrt{-g} \left[ \frac{M_P^2}{2}R + \frac{\dot{\phi}^2}{2M^2}G^{00}\right].
\end{align}
Since the Ricci scalar and the Einstein tensor are expanded as
\begin{align}
R &= 6(2H^2 + \dot{H}) + \frac{1}{4}\left[\left(\dot{h}_{ij}\right)^2 - a^{-2}(\partial_{l}h_{ij})^2\right], \\
G^{00} &= 3H^2 - \frac{1}{8}\left[\left(\dot{h}_{ij}\right)^2 + a^{-2}(\partial_{l}h_{ij})^2 \right],
\end{align}
up to second order in gravitons, we get the following action:
\begin{align}
S_{\rm{grav}} 
&= \frac{M_P^2}{8} \int d^{4}x \; a^{3}\left[\left(1-\frac{1}{2}\frac{\dot{\phi}^2}{M_P^2M^2}\right)\left(\dot{h}_{ij}\right)^2
- \frac{1}{a^2}\left(1+\frac{1}{2}\frac{\dot{\phi}^2}{M_P^2M^2}\right)\left(\partial_{l}h_{ij}\right)^2\right],
\label{eq:gravaction}
\end{align}
where we keep only quadratic terms in gravitons. Note that due to the non-minimal derivative coupling, 
the graviton is non-minimally coupled to the oscillating scalar field. In particular, the sound speed
of the graviton is superluminal. However, due to Eq.~\eqref{eq:expansion}, the graviton 
does not have the ghost-instability and the gradient-instability.

\section{Particle production}
\label{sec:particle}
\setcounter{equation}{0}
In the present models, in particular in Case~B, the Hubble parameter has a large oscillating part.
In Ref.~\cite{Ema:2015dka}, it is shown that the oscillation of the Hubble parameter 
causes production of scalar particles and the graviton. 
This is because all particles are coupled to the scale factor if they are not Weyl-invariant.
Thus, in this section, we investigate particle production due to the oscillation of the Hubble parameter.
Moreover, the non-minimal derivative coupling introduces a direct coupling between 
the inflaton $\phi$ and the graviton. We also consider the graviton production due to this direct coupling.
We do not consider the production of fermions and vector bosons here because their couplings to 
the Hubble parameter vanish in the massless limit and hence are suppressed by small mass parameters.

After summarizing general features on particle production, we 
discuss production of scalar particles and the graviton for both Cases~A and B.
We estimate the particle production rate assuming that the coherently oscillating inflaton $\phi(t)$
dominates the universe. However, as already mentioned, Case~B has an instability 
associated with the negative sound speed squared which 
may invalidate this assumption. This fact should be kept in mind when considering the particle production
for Case~B.

\subsection{General discussion on particle production}

Before going to the discussion of the particle production in the present model, 
let us summarize basic properties of the particle production due to a coherently oscillating scalar field $\phi$
with the mass $m_{\rm eff}$~ \cite{Dolgov:1989us,Traschen:1990sw,Kofman:1994rk}.
We call the produced (bosonic) particle as $\chi$ here.
In general, when a field couples to the coherently oscillating scalar field, the mass of the field also oscillates
because of that interaction. 
Due to the periodic dispersion relation, it is well known that
the coupled field can be copiously produced~\cite{Kofman:1994rk}, whose behavior can be
seen in the stability/instability chart, for instance, of Mathieu equation
for the sine or cosine oscillation.
However, in the following,
we will not dig into technical details of the equation under the periodic background.
Instead, we demonstrate that its typical behavior can be understood more intuitively.

We define a resonance parameter $q$ as $|\Delta m_\chi^{2}| \equiv qm_{\rm{eff}}^{2}$, 
where $\Delta m_\chi^{2}$ is an oscillating part of $\chi$'s mass squared due to 
the $\phi$-dependence and $|\Delta m_\chi^{2}|$ is its amplitude.
Then, the interaction term can be written as\footnote{
For interactions of the type with even powers of $\phi$, such as $\lambda \phi^{2}\chi^{2}$, it may be better to view the process as an annihilation, i.e. 
$\mathcal{L}_{\rm{int}} \sim (qm_{\rm{eff}}/\Phi_{c}^{2})\phi_{c}^{2}\chi^{2}$.
The estimation below, however, remains intact after interpreting the energy of $\chi$ produced by the decay (annihilation)
as $m_{\rm eff}/2$ $(m_{\rm eff})$. }
\begin{align}
\mathcal{L}_{\rm{int}} &\sim \Delta m_{\chi}^{2} \chi_c^{2} \nonumber \\
&\sim  \frac{qm_{\rm{eff}}^{2}}{\Phi_{c}}\phi_{c}\chi_c^{2}  ,
\end{align}
where $\phi_{c}$ and $\chi_c$ are the canonically normalized fields, which may be different from
the original $\phi$ and $\chi$ by an overall factor.\footnote{
In the case with the non-minimal derivative coupling, $\phi_{c} \sim H\phi/M$ for $H/M \gg 1$, 
and $\phi_{c} = \phi$ for $H/M \ll 1$.
}
In addition, $\Phi_{c}$ is the amplitude of $\phi_{c}$.
From this interaction, we can easily estimate the perturbative decay rate of $\phi$ to $\chi$ as
\begin{align}
\Gamma_{\phi} 
\simeq \frac{q^{2}m_{\rm{eff}}^{3}}{8\pi\Phi_{c}^{2}}.
\label{eq:decay_rate}
\end{align}
Note that we have assumed $q \lesssim 1$:
for $q\lesssim 1$, the ``decay'' of $\phi$ into $\chi$ particles is kinematically allowed, and hence the above estimation is justified.
Otherwise, more elaborated treatment of the particle production is needed~\cite{Kofman:1994rk}.
In this paper, we will only encounter the situation with $q\lesssim 1$.

Up to here,
our arguments do not depend on whether or not
the resonance can take place.
In the following, we focus on the narrow resonance regime ($q \lesssim 1$),
which originates from the induced emission of previously produced $\chi$-particles.
If the resonance occurs,
it is well-known that there are resonance bands depending on the momentum 
$k$ of $\chi$ \cite{Kofman:1994rk}. 
In the resonance bands, the wave function of $\chi$ experiences exponential 
growth $\propto e^{\mu t}$ due to the coherent oscillation of $\phi$. The growth rate $\mu$ can also be
related to $q$ as follows.
The time evolution of the number density of $\chi$ for 
$t\gtrsim 1/(qm_{\rm eff})$ is given as\footnote{
At first, for $f_\chi \ll 1$, the distribution function $f_\chi$ has relatively broad width of $\Delta k/k \sim 1/(m_{\rm eff} t)$
due to the uncertainty principle and $f_\chi$ grows as $\propto t^2$, 
but the number density of $\chi$ follows the ordinary evolution equation~\cite{Asaka:2010kv},
\begin{align}
	\dot n_\chi \sim n_\phi \Gamma_\phi.
\end{align}
After $t\gtrsim 1/(qm_{\rm eff})$, the band width becomes $\Delta k\sim qm_{\rm eff}$.
At the same time we have $f_\chi \sim 1$, and after that,
the induced emission becomes significant.
Hence, we can use \eqref{eq:number_evol} for $t\gtrsim 1/(qm_{\rm eff})$.
}
\begin{align}
\dot{n}_{\chi} 
\sim n_{\phi}\Gamma_{\phi} f_{\chi},
\label{eq:number_evol}
\end{align}
where $n_{\chi}$ and $n_{\phi} \equiv \rho_{\phi}/m_{\rm{eff}}$ are the number density of $\chi$ and $\phi$, respectively.
For the coherently oscillating scalar field $\phi$, the number density is estimated as
\begin{align}
n_{\phi} &\sim \frac{V}{m_{\rm{eff}}} \sim m_{\rm{eff}}\Phi^{2}_{c},
\label{eq:number_phi}
\end{align}
where we assumed
$V \sim m_{\rm{eff}}^{2}\Phi_{c}^{2}$.\footnote{
In our case, this assumption is justified both for $H/M \gg 1$ and $H/M \ll 1$.
}
If we define the momentum width of the resonance band as $\Delta k$, the distribution function can be estimated as
\begin{align}
f_{\chi} \sim \frac{2\pi^2 n_{\chi}}{k^{2}\Delta k}.
\end{align}
Since the band width arises from the oscillation of the mass $qm_{\rm{eff}}^{2}$, $\Delta k$ is estimated from
$k \Delta k \sim qm_{\rm{eff}}^{2}$. 
Also, the momentum at the resonance band is given as $k \sim m_{\rm{eff}}$.
Note that the momentum width from the mass oscillation 
must dominate over that from the decay width of $\phi$,
\begin{align}
	q m_\text{eff} > \Gamma_\phi,
\end{align}
for the resonance to take place.\footnote{
	We can prove this as follows. Assuming that $\phi$ completely decays into $\chi$ and $\Gamma_\phi > qm_{\rm eff}$, the maximum phase space density of $\chi$
	is given by $f_\chi^{\rm (max)} \sim n_\phi/(k^2\Delta k) \sim \Phi_c^2/(m_{\rm eff}\Gamma_\phi)\sim q^{2}m_{\rm{eff}}^{2}/\Gamma_{\phi}^{2}$.
	Thus we obtain $f_\chi^{\rm (max)} \lesssim 1$, which means that the resonance does not happen for $\Gamma_\phi > qm_{\rm eff}$.
}
Collecting all these ingredients and Eq.~\eqref{eq:decay_rate} together, 
we can estimate the growth rate $\mu$ from Eq.~\eqref{eq:number_evol} as
\begin{align}
\mu \equiv \frac{\dot{n}_{\chi}}{n_{\chi}} \sim qm_{\rm{eff}}.
\label{eq:growth_rate}
\end{align}
This implies that the number density of $\chi$ exponentially grows as $n_\chi \propto e^{\mu t}$.
This is why we call $q$ the resonance parameter from the first line.
We would like to stress here again that technical details are not required to understand these properties.

In the expanding universe, in order for the resonance to occur, the wave function must be enhanced enough
before their momenta redshift out from the resonance band. Typical timescale for the momentum within the band $\Delta k$
escapes from the band is $\Delta t_H\sim \Delta k / (kH) \sim q/H$.
Requiring that $\Delta t_H$ is longer than the timescale of the resonant enhancement $\sim 1/\mu$, we obtain~\cite{Kofman:1994rk}
\begin{align}
q^{2}m_{\rm{eff}} > H.
\label{eq:resonance_cond}
\end{align}
It is noticeable that this condition ensures $q m_\text{eff} > \Gamma_\phi$
for $q < 1$ and $\Gamma_\phi < H$.
Eqs.~\eqref{eq:decay_rate}, \eqref{eq:growth_rate} and \eqref{eq:resonance_cond} are
extensively used in the discussion in this section.

\subsubsection{Examples}

Here we briefly see the validity of our formula presented above.
We take the potential to be $V(\phi) = m_{\phi}^{2}\phi^{2}/2$.
Let us first consider the following coupling: $\mathcal L = \mu \phi \chi^2$ with canonically normalized scalar field $\phi$ and $\chi$.
For this coupling, we easily obtain $q=\mu \Phi/m_\phi^2$.
From Eq.~(\ref{eq:decay_rate}) the decay rate of $\phi$ is estimated to be $\Gamma_{\phi} \simeq \mu^2/(8\pi m_\phi)$,
which is a well known result.

Next let us consider the following coupling,
\begin{equation}
	\mathcal L = -F^2(\phi)(\partial \chi)^2.
\end{equation}
where $F(\phi)$ is an arbitrary function of $\phi$.
In this case, by defining $\chi_c = F(\phi)\chi$, we obtain the following equation of motion of $\chi_c$,
\begin{align}
	\partial^2 \chi_c - \frac{\partial^2 F}{F}\chi_c = 0.  \label{Eq:chi_c}
\end{align}
In terms of $\chi_c$, it seems that as if it obtains a mass term in the time-dependent background $\phi(t)$.
Thus, for homogeneous $\phi$ field, we obtain $q\simeq \ddot F/(F m_\phi^2) \sim F_{,\phi}\Phi/F$.
Hence, for the simple case of $F = 1+ \phi/\Lambda$ with $|\phi|\ll \Lambda$, we obtain 
$\Gamma_{\phi} \simeq m_\phi^3 /(8\pi \Lambda^2)$
from Eq.~(\ref{eq:decay_rate}),
reproducing the known results of perturbative decay rates.

\subsection{Production of scalar particle}   \label{sec:scalar}

As we have seen in Sec.~\ref{sec:background}, in particular in Case~B, the Hubble parameter has a large oscillating
part in the present model.
In Ref.~\cite{Ema:2015dka}, it is shown that  $\phi$ can decay 
into other non-Weyl invariant particles through the oscillation of the Hubble parameter
since $\phi$ is related to the Hubble parameter via the Friedmann equation. 
Thus, in this subsection we estimate the 
decay rate of $\phi$ to another scalar particle $\chi$ through the oscillation of the Hubble parameter.
We take the mass of $\chi$ to be negligible here.

The action of $\chi$ field is given by
\begin{align}
S_\chi
&= \int d^4x \sqrt{-g}\left(-\frac{1}{2}g^{\mu \nu}\partial_\mu \chi \partial_\nu \chi \right).
\end{align}
We take only the background part of the metric as
\begin{align}
S_\chi
&= \frac{1}{2}\int d^4x \; a^3 \left[ \dot{\chi}^2 - a^{-2}(\partial_i \chi)^2 \right].
\end{align}
Then, as done around Eq.~(\ref{Eq:chi_c}), we can deduce
the resonance parameter $q^{(\chi)}$ as
\begin{align}
q^{(\chi)} \sim \frac{\dot{H}}{m_{\rm{eff}}^{2}},
\end{align}
where we have used the fact that $\dot{H}$ is always larger than or comparable to $H^{2}$.
Thus we estimate the decay rate as
\begin{align}
\Gamma_{\phi\rightarrow\chi} &\sim \frac{\dot H^{2}}{m_{\rm{eff}} \Phi_c^2}.
\end{align}
Now let us consider Cases~A and B separately. 
Note that $\left(q^{(\chi)}\right)^2 m_{\rm eff} < H$ is always satisfied and hence parametric 
resonance does not occur for $\chi$.

\subsubsection{Case~A}

First let us consider Case~A. In this case, we always have $H<M$ and 
the canonically normalized amplitude is the same as the original one: $\Phi_{c} = \Phi$, and $H\sim m_{\rm{eff}}\Phi/M_{P}$.
In this case, we have
\begin{align}
	\Gamma_{\phi\rightarrow\chi} &\sim \frac{\Phi^2 m_{\rm eff}^3}{M_P^4}.
\end{align}
This is the same as production rate in the pure Einstein gravity limit, as it should be.
To conclude, for Case~A, the scalar production rate after inflation is always the same as that of the pure 
Einstein gravity case with minimal kinetic term. Here, note that even in the pure Einstein gravity, 
the Hubble parameter has a small oscillating part which causes the particle production~\cite{Ema:2015dka}.

\subsubsection{Case~B}

Next let us consider Case~B.
In this case, at Phase~1~($H> M$), the canonically normalized amplitude is given as 
$\Phi_{c} = H\Phi/M$, and $m_{\rm{eff}}^{2}\Phi^{2}/M_{P}^{2}M^{2} \sim \mathcal{O}(1)$.
At Phase 2 we have $\Phi_{c} = \Phi$, and $H\sim m_{\rm{eff}}\Phi/M_{P}$.
Thus, by using Eq.~\eqref{eq:dotHubble}, we obtain
\begin{equation}
	\Gamma_{\phi\rightarrow\chi} \sim \left \{ \begin{array}{ll}
	  \displaystyle\frac{M^{2}m_{\rm{eff}}}{\Phi^{2}}& ~~{\rm for}~~M < H,\\
	  \displaystyle\frac{\Phi^4 m_{\rm eff}^7}{M^4M_P^6}& ~~{\rm for}~~M^2/m_{\rm eff} < H < M,\\
	  \displaystyle\frac{\Phi^2 m_{\rm eff}^3}{M_P^4}&~~{\rm for}~~H < M^2/m_{\rm eff}.
	\end{array}
	\right.
\end{equation}
Therefore it reduces to the pure Einstein gravity case with the minimal kinetic term for $H < M^2/m_{\rm eff}$.
Note again that we estimate the production rate assuming that the coherent oscillation of $\phi$ dominates 
the universe, although the instability of the scalar perturbation due to the negative sound speed squared 
may invalidate this assumption.

\subsection{Production of graviton}  \label{sec:graviton}

Now let us focus on the graviton. 
Graviton production has two contributions: one is from the oscillation of the Hubble parameter, 
which results in the approximately same rate as those studied in the previous section, after reinterpreting $\chi$ as the graviton.
The other contribution comes from the non-minimal derivative coupling term, in which 
the graviton and the inflaton directly couple with each other.
Hereafter in this subsection we study the latter contribution.

As we saw in the previous section, the action for the graviton is given by
Eq.~\eqref{eq:gravaction}. 
From now on, we neglect the scale factor $a$ in the action. 
This is because the amount of the graviton production due to the scale factor is expected to be the same as that of 
the scalar particle production, which we have already studied in the previous subsection. 
Thus, we consider the following action:
\begin{align}
S_{\rm{grav}}
&= \frac{1}{2} \int d^{4}x \left[\left(1-\frac{1}{2}\frac{\dot{\phi}^2}{M_P^2M^2}\right)\left(\dot{h}_{ij}\right)^2
- \left(1+\frac{1}{2}\frac{\dot{\phi}^2}{M_P^2M^2}\right)\left(\partial_{l}h_{ij}\right)^2\right],
\label{Sgrav}
\end{align}
where we have rescaled the graviton as $M_Ph_{ij}/2 \rightarrow h_{ij}$.
We further define another time $t^{\prime}$ as
\begin{align}
dt \equiv \frac{f}{g}dt^{\prime},
\end{align}
where
\begin{align}
	f^2\equiv 1-\frac{1}{2}\frac{\dot{\phi}^2}{M_P^2M^2},\\
	g^2\equiv 1+\frac{1}{2}\frac{\dot{\phi}^2}{M_P^2M^2}.
\end{align}
Note that $f^2$ is positive definite (see Eq.~\eqref{eq:expansion}).
By using this time coordinate, the action is rewritten as 
\begin{align}
S_{\rm grav} = \int dt^{\prime}dx^{3} \frac{1}{2}\sqrt{1 - \frac{1}{4}\frac{g^{4}}{f^{4}}\frac{(d\phi/dt^{\prime})^{4}}{M_P^{4}M^{4}}}
\left[\left(\frac{\partial h_{ij}}{\partial t^{\prime}}\right)^2 - (\partial_{l}h_{ij})^2 \right].
\end{align}
Then, as done around Eq.~(\ref{Eq:chi_c}), we can deduce
the resonance parameter $q^{(h)}_{\rm non\mathchar`-min}$ as
\begin{align}
q^{(h)}_{\rm non\mathchar`-min} &\sim \frac{m_{\rm{eff}}^{4}\Phi^{4}}{M_{P}^{4}M^{4}},
\end{align}
where we have taken only the leading term of $\dot{\phi}^{2}/M_{P}^{2}M^{2}$.\footnote{
	In this sense, the difference between $t$ and $t'$ can be neglected.
}
This treatment may be justified even in Phase~1 because of the relation~\eqref{eq:expansion}.
From this $q^{(h)}_{\rm non\mathchar`-min}$ is given by
\begin{equation}
	q^{(h)}_{\rm non\mathchar`-min} \sim \left \{ \begin{array}{ll}
	  \displaystyle 1& ~~{\rm for}~~H > M,\\
	  \displaystyle \frac{H^4}{M^4}& ~~{\rm for}~~H < M.\\
	 \end{array}
	\right.
\end{equation}
Therefore $q^{(h)}_{\rm non\mathchar`-min} \lesssim 1$ is always satisfied.
Note that, as already mentioned, the graviton production also occurs through the oscillation of the Hubble parameter
in the same way as the minimally-coupled scalar field $\chi$.
This contribution to the resonance parameter $q$ for the graviton is denoted by $q^{(h)}_{\rm min} = q^{(\chi)}$.
Thus an effective resonance parameter is given by
\begin{align}
	q^{(h)} = {\rm max}\left\{ q^{(h)}_{\rm min},~~q^{(h)}_{\rm non\mathchar`-min}  \right\}.
\end{align}
Since we have already seen $ q^{(\chi)} < 1$ in all cases, we also have $q^{(h)} \lesssim 1$ in all cases.
Then the decay rate of the inflaton into the graviton is given by
\begin{align}
	\Gamma_{\phi\rightarrow h} &\sim \frac{\left(q^{(h)} \right)^2m_{\rm{eff}}^{3}}{\Phi_{c}^{2}},
\end{align}
In the following, we consider the graviton production in Case~A and B separately.

\subsubsection{Case~A}

In this case, it is easily shown that
\begin{equation}
	\frac{q^{(h)}_{\rm min}}{q^{(h)}_{\rm non\mathchar`-min}} \sim \frac{M^4}{H^2 m_{\rm eff}^2} > 1,
\end{equation}
because $H<M$ and $m_{\rm eff} < M$ are always satisfied in this case.
Therefore, in Case~A, the graviton production rate in the inflaton oscillatory regime is always the same that in
the pure Einstein gravity limit, as it should be
\begin{align}
	\Gamma_{\phi\rightarrow h} &\sim \frac{\Phi^2 m_{\rm eff}^3}{M_P^4}.  \label{graviton:Ein}
\end{align}
Note also that $\left(q^{(h)}_{\rm min} \right)^2 m_{\rm eff} < H$ is also satisfied, hence no parametric resonance happens.

\subsubsection{Case~B}

In Case~B some non-trivial phenomena are likely to happen. In this case we have
\begin{equation}
	q^{(h)} \sim \left \{ \begin{array}{ll}
	  \displaystyle 1 & ~~{\rm for}~~M < H,\\
	  \displaystyle \frac{H^4}{M^4} & ~~{\rm for}~~M^2/m_{\rm eff} < H < M,\\
	  \displaystyle \frac{H^2}{m_{\rm eff}^2}&~~{\rm for}~~H < M^2/m_{\rm eff}.
	\end{array}
	\right.
\end{equation}
In the following, we consider the graviton production in Phase~1 and 2 separately.

First, let us consider Phase~1. In this case, the non-minimal kinetic term dominates over the 
usual one, and hence the canonically normalized scalar field is related to the original one 
as $\Phi_{c} \sim H\Phi/M$. Moreover, note that $m_{\rm{eff}}^{2}\Phi^{2}/M_{P}^{2}M^{2} \sim \mathcal{O}(1)$
is always kept in Phase~1 when $\phi$ dominates the universe; recall the discussion given around Eq.~\eqref{eq:qphase1}.
Therefore, the perturbative decay rate is given as
\begin{align}
\Gamma_{\phi\rightarrow h} \sim \frac{M^{2}}{H^{2}}\frac{m_{\rm{eff}}^{3}}{\Phi^{2}}.
\end{align}
Note that this is an increasing function with time. More importantly, Eq.~\eqref{eq:qphase1} means that
\begin{align}
q^{(h)} \sim \mathcal{O}(1),
\end{align}
is always kept in Phase~1 unless the amplitude $\Phi$ is strongly damped due to the particle production.
Therefore, the resonance condition~\eqref{eq:resonance_cond} is satisfied in Phase~1, and
hence the resonance is expected to be induced as long as $\phi(t)$ can be regarded as homogeneous background.
In such a case, the perturbative decay picture is no longer useful, and a careful treatment is needed.
It would lead to the exponential growth of the graviton with wavenumber around $\sim m_{\rm eff}$.
The production could be so efficient that universe may be dominated by gravitons,
leading to the decrease of the inflaton amplitude $\Phi$ so that it saturates the resonance condition~\eqref{eq:resonance_cond}.
These arguments are based on the assumption that $\phi(t)$ is regarded as homogeneous at least for the time scale
of the growth of the graviton $\sim \left(q^{(h)}m_{\rm eff}\right)^{-1}$.
However, this assumption might be violated due to the instability associated with the negative sound speed squared.
We do not go into details of this case.

Next, let us consider Phase~2 with $M^2/m_{\rm eff} < H < M$. In this case, the canonically normalized field $\phi_{c}$ is 
identical to the original field $\phi$. Therefore, the perturbative decay rate is given by
\begin{align}
\Gamma_{\phi\rightarrow h} &\sim \frac{m_{\rm{eff}}^{11}\Phi^{6}}{M_{P}^{8}M^{8}}.
\end{align}
Note that $q^{(h)}$ is a rapidly decreasing function with time in Phase~2. Therefore, 
the resonance does not last long after entering Phase 2.

Here is a comment. We obtained $q^{(h)} \propto \Phi^{4}$ and $\Gamma_{\phi\rightarrow h} \propto \Phi^6$ 
(except for the $\Phi$ dependence of $m_{\rm eff}$)
although the coupling between $\phi$ and the graviton in the original action~\eqref{eq:gravaction} is of the form of $\phi^{2}h^{2}$, 
from which one may naively expect $q^{(h)} \propto \Phi^2$ and $\Gamma_{\phi\rightarrow h} \propto \Phi^2$. 
This property may be understood in the language of the original action~\eqref{eq:gravaction} as follows. 
First, we must note that the coupling in Eq.~\eqref{eq:gravaction} is in the following form:
\begin{align}
\mathcal{L}_{\rm{int}} 
	&\sim \frac{\dot{\phi}^2}{M_P^2M^2}\left[\left(\dot{h}_{ij}\right)^2 + (\partial_{l}h_{ij})^2\right].
\end{align}
If we consider decay of $\phi$ particle at rest into two gravitons, the amplitude would be proportional to $E^2 - |\vec{p}|^2$
where $E$ and $\vec{p}$ are the energy and momentum of the produced gravitons, and we have used 
the 4-momentum conservation. For the usual on-shell gravitons, $E^2 - |\vec{p}|^2$ would vanish
since the graviton is massless. In the present case, however, the dispersion relation of the graviton is also modified 
due to the non-minimal derivative coupling, and hence we get
\begin{align}
E^2 - |\vec{p}|^2 \sim \frac{m_{\rm{eff}}^4\Phi^2}{M_P^2M^2}.
\end{align}
Therefore, the decay width is expected to be proportional to $\Phi^6$.
Again note that we simply neglect the effects of the instability here.

Finally, for $H < M^2/m_{\rm eff}$, the graviton production rate becomes the same as the pure Einstein limit \eqref{graviton:Ein}.

\section{Conclusion}
\label{sec:conclusion}
\setcounter{equation}{0}

In this paper, we have investigated the particle production in the inflationary models with the non-minimal derivative
coupling $G^{\mu \nu}\partial_\mu \phi \partial_\nu \phi$ between the inflaton and the Einstein tensor, 
focusing especially on the inflaton oscillation regime.
In this model, when the non-minimal part dominates the inflaton kinetic term (Phase~1), the Hubble parameter violently 
oscillates with time if the inflaton oscillation dominates the universe. The oscillation of the Hubble parameter
causes particle production, and we have estimated the decay rate of the inflaton in that process.
More importantly, the non-minimal derivative coupling introduces a direct coupling between the inflaton and the 
graviton. 
We have seen that, via the non-minimal derivative coupling, 
the resonant graviton production may occur due to the inflaton oscillation in Phase~1. 
As a result, the energy density of the inflaton may be efficiently transferred into that of the graviton, and
the graviton may dominate the universe.

However, in Phase 1, the sound speed squared of the inflaton rapidly oscillates between positive and negative values.
It may result in the strong instability for the shortest wavelength mode.
It may or may not be phenomenologically problematic, but analysis in such a violent situation is beyond the scope of this paper.
A conclusion is that if we simply avoid such an instability within this effective field theory, the particle production rates are the same as
those in the models with Einstein gravity with minimal kinetic term as found in Ref.~\cite{Ema:2015dka}.
Also it should be noticed that inflation models with non-minimal derivative coupling taking dominant role
such as the new Higgs inflation model~\cite{Germani:2010gm}
necessarily experience this instability after inflation. 

Although we have assumed in this paper that the non-minimal coupling between the inflaton and the Einstein tensor 
takes the form ${\mathcal L} \sim G^{\mu \nu} \partial_\mu \phi \partial_\nu \phi$,
the graviton production is expected to occur in more general $G_{\mu\nu}$-type models ($G_{5}$-terms
in the context of the generalized G-inflation\cite{Kobayashi:2011nu}). This is because a direct coupling between the inflaton
and graviton is introduced in these models, and hence the graviton is expected to be
produced efficiently, as in the case of the present model.
However, we should also be careful on the instability of perturbations in these models.

\section*{Acknowledgments}

This work was supported by the Grant-in-Aid for Scientific Research on Scientific Research A (No.26247042 [KN]),
Young Scientists B (No.26800121 [KN]) and Innovative Areas (No.26104009 [KN]).
This work was supported by World Premier International Research Center Initiative (WPI Initiative), MEXT, Japan. 
The work of Y.E., R.J. and K.M. was supported in part by JSPS Research Fellowships for Young Scientists.
The work of Y.E. and R.J. was also supported in part by the Program for Leading Graduate Schools, MEXT, Japan.

\appendix

\section{Inflation with non-minimal derivative coupling}
\label{app:perturbation}
\setcounter{equation}{0}

In this Appendix we briefly discuss slow-roll inflation when the non-minimal derivative part of the kinetic terms takes a dominant role.
Since $G^{00}\simeq 3H^2$ and $G^{ij}\simeq-3H^2/a^2$ during inflation, the Lagrangian of inflaton $\phi$ is given as\footnote{
	Precisely speaking, $G^{00} \neq -a^2G^{ij}$ and hence the sound speed of $\delta\phi$ is modified.
	But its effect is small as long as we are concerning physical quantities up to the first order in slow-roll parameters.
}
\begin{equation}
	\mathcal L \simeq -\frac{3H^2}{2M^2}(\partial\phi)^2- V(\phi) \simeq -\frac{\lambda \phi^n}{2nM^2M_P^2}(\partial\phi)^2- V(\phi).
\end{equation}
We can define canonically normalized field
\begin{equation}
	\tilde\phi \equiv \frac{\sqrt{\lambda}}{\sqrt{n}MM_P}\phi^{(n+2)/2},
\end{equation}
and the Lagrangian is rewritten as
\begin{equation}
	\mathcal L \simeq -\frac{1}{2}(\partial\tilde\phi)^2 - V(\tilde\phi),~~~
	V(\tilde\phi)=\left( \frac{\lambda}{n} \right)^{\frac{2}{n+2}} \left(MM_P\tilde\phi \right)^{\frac{2n}{n+2}}.
\end{equation}
Note that this transformation is meaningful only in the slow-roll regime.
After inflation, we cannot approximate as $3M_P^2 H^2 \simeq V$, hence such a transformation is not possible.
This is the difference from the running kinetic inflation~\cite{Takahashi:2010ky}.

Therefore, in terms of $\tilde\phi$, the potential just behaves like $\propto \tilde\phi^{2n/(n+2)}$.
The slow-roll equation of motion is
\begin{equation}
	3H\dot{\tilde\phi} + V_{\tilde\phi}=0,~~~~~V_{\tilde\phi}\equiv \frac{\partial V(\tilde\phi)}{\partial \tilde\phi}.
\end{equation}
According to the standard procedure, slow-roll parameters at the e-folding number $N$ is calculated as
\begin{gather}
	\epsilon_V = \frac{1}{2}M_P^2\left( \frac{V_{\tilde\phi}}{V} \right)^2=\frac{n}{2(n+2)N},~~~~~~
	\eta_V = M_P^2 \frac{V_{\tilde\phi \tilde\phi}}{V} = \frac{n-2}{2(n+2)N}.
\end{gather}
The slow-roll conditions $\epsilon_V, |\eta_V| \ll 1$ are equivalent to $m_{\rm eff} \ll H$.
We obtain the scalar spectral index and the tensor-to-scalar ratio as
\begin{gather}
	n_s = 1-6\epsilon_V + 2\eta_V = 1-\frac{2(n+1)}{(n+2)N},~~~~~~
	r = 16\epsilon_V = \frac{8n}{(n+2)N}.
\end{gather}
These are inside the 2$\sigma$ range of the Planck constraint for $n\leq 4$~\cite{Tsujikawa:2013ila,Germani:2011ua,Ade:2015lrj}.
In order to fit the CMB normalization $\mathcal P_\zeta =2.2\times 10^{-9}$ observed by Planck~\cite{Ade:2015lrj}, we need
\begin{equation}
	\frac{m_\phi M}{M_P^2} \simeq 2\times 10^{-10}\left( \frac{50}{N} \right)^{3/2}~~~{\rm for}~~~n=2  \label{CMB_n2}
\end{equation}
and
\begin{equation}
	\frac{\lambda M^4}{M_P^4} \simeq 1\times 10^{-31} \left( \frac{50}{N} \right)^5 ~~~{\rm for}~~~n=4.  \label{CMB_n4}
\end{equation}

Note that in order for inflation to take place in the non-minimal regime, we need the following condition:
\begin{equation}
	\left.\frac{3H^2}{M^2}\right|_{\phi=\phi_e} \simeq \left( \frac{\lambda M_P^{n-2}}{M^2} \right)^{\frac{2}{n+2}}  \gg 1,
	\label{nonmin_cond}
\end{equation}
where $\phi_e$ is the field value at the end of inflation.
This implies $m_\phi \gg M$ for $n=2$ and $\lambda \gg (M/M_P)^2$ for $n=4$.
Thus we can take $m_\phi \gg 10^{13}\,$GeV for $n=2$ and $\lambda \gg 10^{-10}$ for $n=4$ 
by choosing $M$ appropriately while keeping the CMB normalization consistent with observations.

\section{Proof of adiabaticity}
\label{app:adiabaticity}
\setcounter{equation}{0}
In this Appendix, we prove the adiabaticity of $J$.
The energy conservation Eq.~\eqref{eq:rhodot} reads
\begin{align}
\frac{d}{dt}\left[\left(1+\frac{9H^2}{M^2}\right)\frac{\dot{\phi}^2}{2} + V \right] 
+ 3H\left(1+\frac{3H^2}{M^2}\right)\dot{\phi}^2 - \frac{3H}{M^2}\frac{d}{dt}\left(H\dot{\phi}^2\right)
&= 0.
\end{align}
The second term is of the order of $\sim \mathcal{O}(H\rho_{\phi})$, and hence we safely
neglect this term here. Thus,
we must find a new adiabatic invariant from the equation
\begin{align}
\frac{d}{dt}\left[\left(1+\frac{9H^2}{M^2}\right)\frac{\dot{\phi}^2}{2} + V \right] 
	 - \frac{3H}{M^2}\frac{d}{dt}\left(H\dot{\phi}^2\right) \simeq 0.
\end{align}
In order to find an adiabatic invariant in this system, we define the following quantity $I$:
\begin{align}
I \equiv \frac{1}{H^{c_{1}}}\left[\left(1+\frac{c_{2}H^2}{M^2}\right)\frac{\dot{\phi}^2}{2} + V \right],
\end{align}
where $c_{1}$ and $c_{2}$ are some numbers we determine below. By taking derivative with respect to time, 
we get
\begin{align}
H^{c_{1}}\frac{dI}{dt} 
	=& \frac{d}{dt}\left[\left(1+\frac{9H^2}{M^2}\right)\frac{\dot{\phi}^2}{2}+V\right]
	- \frac{9-c_{2}}{2} \frac{H}{M^2}\frac{d}{dt}\left(H\dot{\phi}^2\right) \nonumber \\
	&-\left(c_{1}c_{2} - 9c_{1} - c_{2} + 9\right)\frac{\dot{\phi}^2}{2M^2}H\dot{H} - 3M_P^2c_{1}H\dot{H},
\end{align}
where we have used $\dot{\phi}^2/2 + V = 3M_P^2H^2(1 - 3\dot{\phi}^2/2M_P^2M^2)$.
Therefore, for
\begin{align}
c_{1} = 1, \\
c_{2} = 3,
\end{align}
the following equation is obtained:
\begin{align}
\frac{d}{dt}\left(I + 3M_P^2H\right) \simeq 0.
\end{align}
This means that the quantity $I+3M_P^2H$ is adiabatically conserved.
This quantity is related to $J$ as 
\begin{align}
2J = I + 3M_P^2H.
\end{align}
Therefore, we have proved that $J$ is an adiabatic invariant.
Note that the adiabaticity of $J$ holds for all Phases~0--2.
A more systematic way to derive an adiabatic invariant for general actions is described in~\cite{Ema:2015eqa}.

\section{Sound speed squared in another gauge}
\label{app:delphi_gauge}
\setcounter{equation}{0}
In this appendix, we show that the formula of the sound speed squared of the scalar perturbation~\eqref{eq:sound} does not 
change in another gauge which is well defined even at the end points of the oscillation, $\dot{\phi} = 0$. 
We again use the ADM formalism, where the metric is given as
\begin{align}
ds^{2} = -N^{2}dt^{2} + \gamma_{ij}\left(dx^{i} + \beta^{i}dt\right)\left(dx^{j}+\beta^{j}dt\right).
\end{align}
Here we take the gauge condition as
\begin{align}
	N &= 1+\alpha, \nonumber \\
	\beta_{i} &= \partial_{i}\psi, \nonumber \\
	\gamma_{ij} &= a(t)^{2}\delta_{ij}, \nonumber \\
	\phi &= \bar{\phi}(t) + \delta \phi,
	\label{eq:other_gauge}
\end{align}
where we have neglected the vector and tensor modes since they do not affect the quadratic action
of the scalar perturbation. Here $\bar{\phi}(t)$ is the background part of the scalar field in this gauge.
Note that we can take this gauge condition even at $\dot{\bar{\phi}} = 0$ points.
We treat $\alpha, \psi$ and $\delta \phi$ as perturbations and expand the action with respect to them.
Then, the action is given to the second order as follows:
\begin{align}
S = \int d^4x\; a^3 &\left[\frac{1}{2}\left(1+\frac{3H^2}{M^{2}}\right)\delta \dot{\phi}^2 
- \frac{1}{2a^2}\left(1+\frac{3H^{2}}{M^{2}}+\frac{2\dot{H}}{M^{2}}\right)\left(\partial_{i}\delta \phi\right)^2
-\frac{1}{2}V^{\prime\prime}\delta \phi^2 \right. \nonumber \\
&- \left. \dot{\bar{\phi}}\left(1+\frac{9H^2}{M^2}\right)\alpha \delta \dot{\phi} 
- \frac{2H\dot{\bar{\phi}}}{a^2M^2}\partial_{i}\alpha \partial_{i}\delta \phi - V^{\prime}\alpha \delta\phi \right. \nonumber \\
&+ \left. \left(-3M_{P}^2H^2 + \frac{\dot{\bar{\phi}}^{2}}{2}\left(1+\frac{18H^{2}}{M^2}\right)\right)\alpha^2 \right. \nonumber \\
&+ \left. \frac{1}{a^{2}}\left(\dot{\bar{\phi}}\left(1+\frac{3H^{2}}{M^{2}}\right)\delta \phi 
- \frac{2H\dot{\bar{\phi}}}{M^2}\delta \dot{\phi} 
-2HM_{P}^2 \left(1-\frac{3}{2}\frac{\dot{\bar{\phi}}^2}{M^2M_{P}^2}\right)\alpha \right)\partial^2 \psi \right],
\label{eq:action_delphi}
\end{align}
where we have omitted the background part of the action. Also, first order terms vanish due to the 
background equation of motion. 

Two important features can be read from the above action. 
Firstly, $\alpha$ and $\psi$ does not have the kinetic term, and hence their equations of motion
give constraint equations among $\delta \phi, \alpha$ and $\psi$. Secondly, the action depends only 
linearly on $\psi$. This means that, after solving the equation of motion, $\psi$ does not contribute
to the original action. Thus, we need not explicitly solve the equation of motion for $\alpha$ to get 
the form of $\psi$ in terms of $\delta \phi$. This feature greatly simplifies the calculation. 

By solving the equation
of motion for $\psi$, we obtain the following relation:
\begin{align}
\alpha = \frac{\dot{\bar{\phi}}}{2HM_{P}^{2}\left(1-\frac{3}{2}\epsilon\right)}
\left[\left(1+\frac{3H^2}{M^2}\right)\delta \phi - \frac{2H}{M^2}\delta \dot{\phi}\right],
\end{align}
where $\epsilon$ is defined in Eq.~\eqref{eq:expansion}. After substituting this into Eq.~\eqref{eq:action_delphi} and
doing some integration by parts, we obtain the following kinetic term for $\delta \phi$:
\begin{align}
S_{\rm{kin}} = M_{P}^{2}M^2\int dx^{4}\; a^{3}\frac{F^{2}G}{\dot{\bar{\phi}}^2}
\left[\delta \dot{\phi}^{2} - \frac{c_{s}^{2}}{a^{2}}\left(\partial_{i}\delta \phi\right)^{2}\right],
\label{eq:quad_delphi}
\end{align}
where the definitions of $F, G$ and $c_{s}^2$ are the same as those given in Sec.~\ref{sec:scalarpert}, 
and we have omitted the mass term for $\delta \phi$ since it is not relevant to the sound speed squared.
In order to obtain this expression, the following background equation
\begin{align}
-\frac{\dot{H}}{M^2}\left(1-\frac{\epsilon}{2}\right) = \frac{1}{2}\left(1+\frac{3H^2}{M^2}\right)\epsilon - \frac{H}{2M^2}\dot{\epsilon},
\end{align}
is useful. Note that the action~\eqref{eq:quad_delphi} is well defined even at $\dot{\bar{\phi}} = 0$ points
since $G \propto \dot{\bar{\phi}}^2$. Thus, we conclude that the formula of the sound speed squared 
is the same in the other gauge condition~\eqref{eq:other_gauge}. Note that, after including
the mass term, the actions~\eqref{eq:quad-action} and \eqref{eq:quad_delphi} are equivalent once we identify 
$\zeta$ and $\delta \phi$ as
\begin{align}
\zeta = -\frac{H}{\dot{\bar{\phi}}}\delta \phi.
\end{align}
This is nothing but the gauge transformation from Eq.~\eqref{eq:other_gauge} to Eq.~\eqref{eq:gauge}.



\end{document}